\documentclass[10pt, a4paper]{article}
\usepackage{booktabs}
\usepackage{multirow}
\usepackage{url}
\usepackage{tcolorbox}
\usepackage{tipa}

\def\BibTeX{{\rm B\kern-.05em{\sc i\kern-.025em b}\kern-.08em
    T\kern-.1667em\lower.7ex\hbox{E}\kern-.125emX}}
\usepackage[final]{lrec2026} % this is the new style
\title{The PARLO Dementia Corpus: \\ A German Multi-Center Resource for Alzheimer's Disease}

\name{
Franziska Braun$^{\ast}$,
Christopher Witzl$^{\ast}$,
Florian Hönig$^{\dagger}$, \\
{\bf \large Elmar Nöth$^{\ddagger}$,
Tobias Bocklet$^{\ast}$,
Korbinian Riedhammer$^{\ast}$}
}

\address{$^{\ast}$Technische Hochschule Nürnberg, Germany \\ $^{\dagger}$KST Institut GmbH Bad Emstal, Germany \\ $^{\ddagger}$Friedrich-Alexander-Universität Erlangen, Germany\\
franziska.braun@th-nuernberg.de, korbinian.riedhammer@th-nuernberg.de\\}

\abstract{
Early and accessible detection of Alzheimer's disease (AD) remains a major challenge, as current diagnostic methods often rely on costly and invasive biomarkers. 
Speech and language analysis has emerged as a promising non-invasive and scalable approach to detecting cognitive impairment, but research in this area is hindered by the lack of publicly available datasets, especially for languages other than English. 
This paper introduces the PARLO Dementia Corpus (PDC), a new multi-center, clinically validated German resource for AD collected across nine academic memory clinics in Germany. 
The dataset comprises speech recordings from individuals with AD-related mild cognitive impairment and mild to moderate dementia, as well as cognitively healthy controls. 
Speech was elicited using a standardized test battery of eight neuropsychological tasks, including confrontation naming, verbal fluency, word repetition, picture description, story reading, and recall tasks. 
In addition to audio recordings, the dataset includes manually verified transcriptions and detailed demographic, clinical, and biomarker metadata.
Baseline experiments on ASR benchmarking, automated test evaluation, and LLM-based classification illustrate the feasibility of automatic, speech-based cognitive assessment and highlight the diagnostic value of recall-driven speech production.
The PDC thus establishes the first publicly available German benchmark for multi-modal and cross-lingual research on neurodegenerative diseases.
\\ \newline \Keywords{dementia screening, pathological speech, neuropsychological tests} 
}

\begin{document}

\maketitleabstract

\section{Introduction}
Alzheimer's disease (AD) is the most prevalent cause of dementia worldwide and poses major challenges to healthcare systems in aging societies. 
The progressive decline in memory, language and executive functions profoundly affects the autonomy and quality of life of those affected. 
Early detection is essential to enable timely intervention, treatment planning, and inclusion in clinical trials for emerging disease-modifying therapies.
However, current diagnostic procedures rely largely on costly and invasive biomarkers such as Positron Emission Tomography (PET) imaging or cerebrospinal fluid analysis, which are not available in many clinical or community settings.

In recent years, speech and language analysis has gained increasing attention as a non-invasive, cost-effective, and scalable approach to detecting cognitive impairment. 
A growing body of work shows that acoustic and linguistic features extracted from spontaneous or semi-structured speech can serve as early indicators of mild cognitive impairment and dementia \cite{ramanarayanan2024multimodal}. 
Speech is a particularly appealing biomarker, as it can be recorded remotely, repeatedly, and unobtrusively using consumer devices.
Despite rapid methodological advances, speech resources are still very limited. 
Most publicly available corpora, such as the Pitt Corpus (DementiaBank) \cite{becker1994dementiabank}, are small, English-centric, and often limited to single elicitation tasks (e.g., picture description) as opposed to standardized tests. 
As a result, cross-lingual generalization and reproducibility remain limited. 
In particular, German resources for investigating cognitive impairment based on speech are rare and not publicly available, and only a few datasets combine clinically verified diagnoses with standardized neuropsychological tasks and human transcriptions.

This paper introduces the PARLO Dementia Corpus (PDC), a new German multi-center corpus for Alzheimer's disease, which was collected at nine academic memory clinics across Germany as part of a study conducted by the PARLO Institute for Research and Teaching in Speech Therapy. 
The dataset includes recordings from 208 German-speaking participants, including AD patients with mild cognitive impairment (MCI) and mild to moderate Dementia (DEM), as well as cognitively healthy controls (HC) of comparable age. 
Speech data was elicited using a standardized iPad-based test battery with eight cognitive tasks, including story reading, verbal fluency (animal naming), confrontation naming, word repetition, picture description, and recall tasks. 
All recordings were collected under uniform clinical conditions and manually transcribed according to the extended scientific transcription rules of Dresing \& Pehl \cite{dresing_pehl10}.
The resulting multi-modal resource includes audio recordings and manually verified transcriptions, enriched with detailed demographic and clinical metadata, including Mini-Mental State Examination (MMSE) scores, diagnostic labels, and biomarkers from cerebrospinal fluid and medical imaging. 
The PDC's design enables research across multiple domains, from automatic speech recognition (ASR) and speech biomarker extraction to automated dementia assessment, as well as cross-lingual and multi-modal studies on cognitive impairment.
To demonstrate the applicability of the dataset, we present three baseline experiments:  
\begin{itemize}
    \item ASR benchmarking using three open-source systems to evaluate transcription quality under different cognitive and task conditions.  
    \item Automatic test evaluation for the \textit{Confrontation Naming} and \textit{Animal Naming} tasks using rule-based scoring to validate the agreement between ASR-based and manual scores. 
    \item Vision-LLM zero-shot classification of cognitive impairment to explore the potential of generative models for automatic dementia assessment.
\end{itemize}

\section{Related Work and Data}
The automatic detection of Alzheimer's disease based on speech has become an active area of research.
Early systematic reviews and meta-analyses \cite{fuente20_review,martinez21_review,vigo22_review} summarize heterogeneous approaches (e.g., acoustic, linguistic, temporal) and report promising discriminatory performance in small clinical cohorts, but they also point to methodological heterogeneity, limited external validation, and biases in the datasets that hinder clinical implementation.
The publicly available Pitt Corpus (DementiaBank) \cite{becker1994dementiabank} and derived benchmark sets have been widely used for picture description tasks (``Cookie Theft'') and verbal fluency analyses, enabling comparisons between different studies. 
Building on these, the ADReSS, ADReSSo, and TAUKADIAL Challenges \cite{adress20,adresso21,taukadial24} provided acoustically preprocessed, age- and gender-balanced datasets with spontaneous speech and tasks such as MCI and AD classification, regression of cognitive scores (MMSE), and prediction of disease progression. 
%and showed that careful experimental design (speaker-level splitting, covariate balancing) is crucial to avoid overoptimistic results.
Work in this field generally follows two complementary strategies: (1) pipelines that extract acoustic and linguistic features, followed by classical classifiers (e.g., SVM, RF, gradient boosting)\cite{ammar18_speechAD,yuan_disfluencies_2020, calza_linguistic_2021,braun_GoingCookieTheft_2022}; and (2) representation learning approaches that fine-tune large pre-trained audio or text encoders (e.g., wav2vec, transformer-based language models) on AD tasks \cite{yuan_DisfluenciesFineTuningPreTrained_2020,balagopalan_BERTNotBERT_2020,balagopalan21_finetune,pappagari21_interspeech}. 
Recent studies show that both acoustic and linguistic signals contain disease-relevant information and that multi-modal fusion often improves robustness \cite{rohanian20_interspeech,pompili20_interspeech,braun24_interspeech,lin24_multimodal}.
Since English-centric studies do not necessarily generalize to other languages, multilingual studies and challenges \cite{perez-toro_alzheimers_2022,madress22} have investigated cross-lingual approaches and shown that some acoustic markers (e.g., wav2vec) can be transferred between languages, while linguistic markers may require language-aware modeling and evaluation. 
This motivates language-specific data collection and validation, especially for German, where large, balanced corpora with standardized collection are relatively rare and furthermore not publicly available \cite{weiner16_interspeech,braun22_interspeech}.

The PARLO Dementia Corpus fills this gap by providing the first German multi-center, clinically annotated resource for AD, enabling reproducible multi-modal and cross-lingual research on cognitive decline.
Subsets of the corpus have already been used in \cite{pereztoro23_interspeech,braun_classifying_2023,braun24_interspeech}, which provided baselines for speech-based classification of HC, MCI, and DEM.
Cross-corpus results have shown that disease-based classification across other German datasets \cite{braun_classifying_2023}, as well as across English and Spanish datasets \cite{pereztoro23_interspeech} is possible using acoustic, linguistic, and emotional embeddings from several subtests of the PDC (e.g., animal naming, confrontation naming, word repetition, and picture description).
Moreover, \cite{braun24_interspeech} established a baseline for the early detection and monitoring of AD using pause-enriched cross-attention with text and audio data from picture description and animal naming tasks of the PDC.

\section{The PARLO Dementia Corpus}
\begin{table}[!th]
  \centering
    \begin{tabular}{c|ccc}
    \toprule
    \textbf{Group} & \textbf{$n$}         & \textbf{Age}               & \textbf{Gender}   \\
    \midrule
    HC  & 83 & 55-87 (69.0$\pm$7.9) & 32m/51f  \\
    MCI & 59 & 55-85 (70.7$\pm$8.4) & 32m/27f  \\
    DEM & 66 & 55-85 (71.7$\pm$8.8) & 32m/34f  \\
    \midrule
    ALL & 208 & 55-87 (70.4$\pm$8.4) & 96m/112f \\
    \bottomrule
    \end{tabular}
    \caption{Demographics for healthy control (HC), mild cognitive impairment (MCI), and dementia (DEM) groups.}
  \label{tab:demo}
\end{table}
The PARLO Dementia Corpus (PDC) consists of audio recordings from a total of 208 German-speaking subjects (96 men, 112 women) aged between 55 and 87 ($\mu=70.4\pm8.4$). 
% Data collection was coordinated by the PARLO Institute for Research and Teaching in Speech Therapy. 
Subjects were recruited as part of a multi-center, cross-sectional, open-label, controlled, parallel group clinical study in nine academic memory clinics across Germany.
Research was approved by the Institutional Review Boards (IRBs) responsible for each study site; each subject provided informed consent prior to recording.
The demographic distribution of the study population, including healthy controls (HC, $n=83$), and individuals with AD-related mild cognitive impairment (MCI, $n=59$) and dementia (DEM, $n=66$), is given in Table~\ref{tab:demo}.

\subsection{Inclusion and Exclusion Criteria}\label{sc:criteria}
The study population consists of male and female participants aged 55 years or older, including patients diagnosed with Alzheimer's disease, ranging from MCI to mild and moderate dementia, as well as cognitively healthy controls of comparable age. 
All participants must have adequate vision and hearing, fluent German language skills, and sufficient verbal abilities to perform the study tasks. 
Legal competence and written informed consent were required for inclusion. 
Participation in registries was permitted for all study subjects.
For AD patients, an established diagnosis of AD according to the National Institute on Aging-Alzheimer's Association (NIA-AA) criteria \cite{clifford2018_niaaa} was required prior to study enrollment. 
For controls, cognitive test scores must fall within normal limits of the CERAD test battery \cite{morris88} ($\leq$1.2 standard deviations below age-adjusted norms). 
Controls must have at least six years of education, report no subjective cognitive complaints, and female participants must be of non-childbearing potential ($\geq$1 year postmenopausal or surgically sterilized).

Subjects were excluded if they had any medical or psychiatric condition interfering with cognitive functioning, including clinical signs of depression, or any condition prohibiting the use of a tablet-based application. 
Other exclusion criteria include ongoing clinically significant metabolic or systemic disease that could impair memory or speech (e.g., untreated thyroid disease, vitamin deficiency, or chronic kidney and liver disease), cognitive impairment attributable to conditions other than early-phase neurodegenerative disease, Parkinson's disease, history of brain tumor, intracranial lesion, disturbance of cerebrospinal fluid circulation (e.g., normal pressure hydrocephalus), significant head trauma, or brain surgery. 
Further exclusion criteria include major cerebrovascular disease, as indicated by clinical history or neuroimaging, the use of psychoactive medications likely to interfere with cognitive testing (e.g., benzodiazepines, sedatives, or antipsychotics), and participation in any parallel clinical trial or investigation.

\subsection{PARLO Dementia Test Battery}\label{sc:tasks}
\begin{figure}[!th]
  \begin{center}
  \includegraphics[width=\columnwidth]{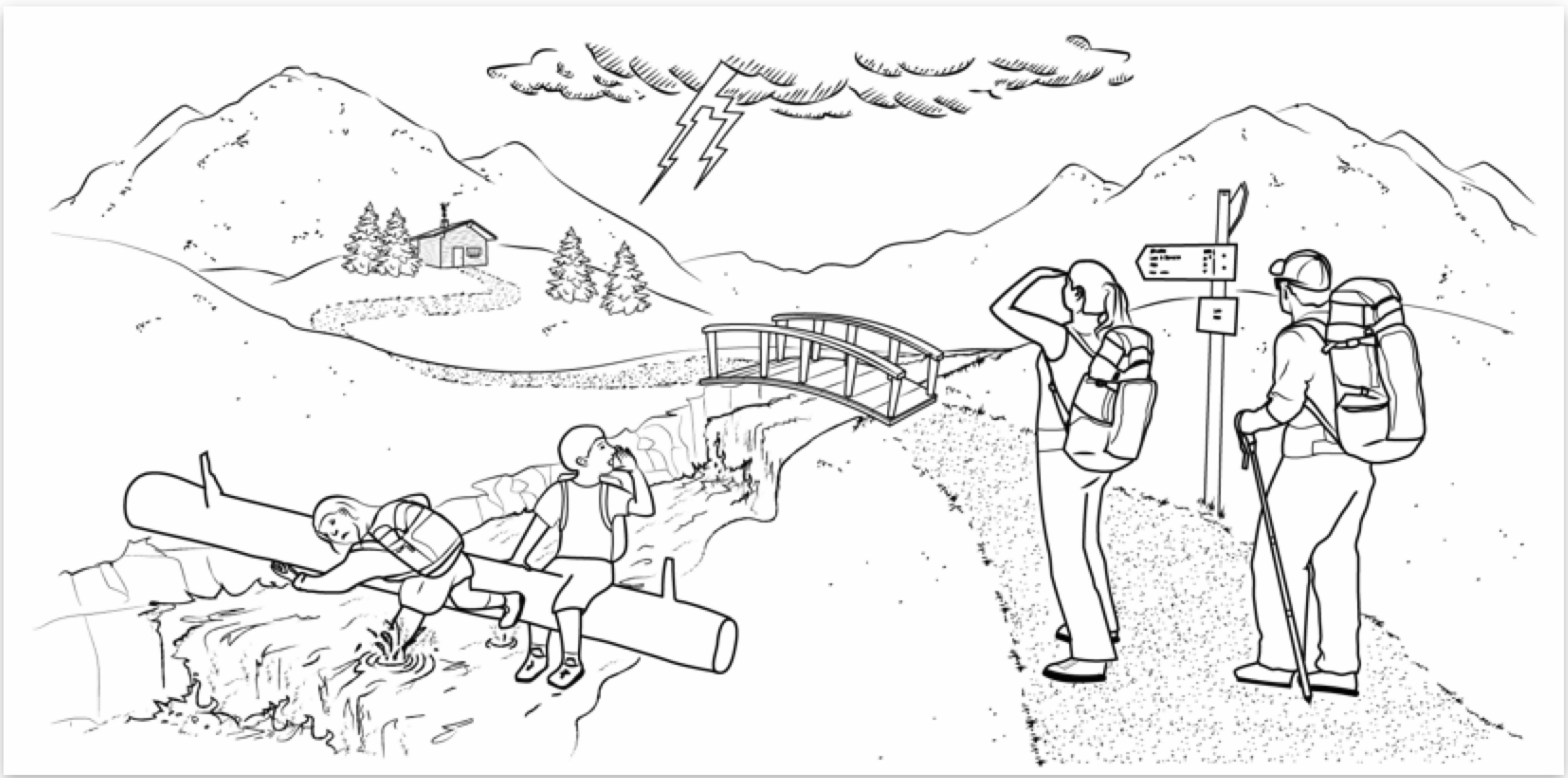}
  \caption{Picture Description Task ``Mountain Scene''}
  \label{fig:mountain}
  \end{center}
\end{figure}
The PARLO Dementia Test Battery (PDTB) comprises eight neuropsychological speech tasks that assess multiple cognitive and linguistic domains, as described below in the order of examination:
\begin{enumerate}
    \item \textbf{Story Reading:} Subjects read aloud a short story (called ``Johanna Subway''). This task ensures uniform stimulus exposure and provides a baseline for prosodic, articulation, and fluency analysis. The ground truth story text is included in the corpus.
    \item \textbf{Confrontation Naming:} Subjects successively view 15 objects (line drawings) and name them. This is a variant of the Boston Naming Test \cite{kaplan78}, which assesses lexical access, confrontation naming, and semantic competence. Deficits in naming can be an indicator of neurodegenerative diseases.
    \item \textbf{Animal Naming:} In one minute, subjects name as many animals as possible. This is a semantic verbal fluency test that assesses lexical retrieval, semantic network integrity, and executive control (switching between subcategories) \cite{isaacs73}. It is widely used in dementia research due to its cross-lingual applicability and sensitivity to AD-related cognitive decline.
    \item \textbf{Picture Description:} Subjects describe the picture of a ``Mountain Scene'' (shown in Figure~\ref{fig:mountain}\footnote{PARLO GmbH, \url{www.parlo-institut.de}}). This elicits spontaneous descriptive discourse, combining visual-to-verbal mapping, semantic content generation, and syntax planning \cite{borod80}. Innovations in speech graph and semantic path modeling can be used to derive features from such descriptions. The original picture is included in the corpus.
    \item \textbf{Word Repetition ``pataka'':} Subjects repeat the pseudo-word ``pataka'' (\textipa{["pataka]}) as often as possible within 10 seconds. This tests phonological processing, speech-motor planning, and phonological loop functioning under minimal lexical interference.
    \item \textbf{Word Repetition ``sischafu'':} Analogous to the ``pataka'' task, subjects repeat another pseudo-word, ``sischafu'' (\textipa{[sISa"fu]}), within 10 seconds. The use of multiple pseudo-words reduces item-specific learning or familiarity, increasing the generalizability of phonological processing measures.
    \item \textbf{Story Recall:} After previous distraction, subjects retell the story in their own words. This tests episodic memory, narrative coherence, and spontaneous language generation, capturing both semantic and syntactic retrieval processes.
    \item \textbf{Picture Recall:} After previous distraction, subjects describe the previously shown picture from memory. This tests episodic memory, reconstruction of spatial and semantic elements, and narrative consistency under memory load.
\end{enumerate}

% Story Reading [sɪʃaˈfu]
% Patients read a short story (Johanna Subway).

% Story Recall
% Patients explain what the story was about in their own words. (Read story loud)

% Animal Naming
% Patients are asked to name as many animals as possible in one minute. (List "Oberbegriff")

% Boston Naming
% Patients name the different objects shown to them on screen. (Name Item visible on picture)

% Word Repetition pataka
% Patients repeat a couple of non-existing words (pataka) during a predefined time of 10 seconds. (Repeat Word as often as possible)

% Word Repetition sischafu
% Patients repeat a couple of non-existing words (sischafu) during a predefined time of 10 seconds. (Repeat Word as often as possible)

% Picture Description
% Patients describe a picture (Mountain Scene) as shown in the app.

% Picture Recall
% Patients describe the same picture as in the previous exercise, this time without the actual picture on screen.

\subsection{Recordings}
\begin{table}[!th]
\begin{center}
\begin{tabular}{l|r|r|r|r}
\toprule
\textbf{Task} & \textbf{ALL} & \textbf{HC} & \textbf{MCI} & \textbf{DEM} \\
\midrule
Confrontation Naming & 204 & 81 & 58 & 65 \\
Animal Naming & 205 & 82 & 58 & 65 \\
\midrule
\textbf{Subset Naming} & 204 & 81 & 58 & 65  \\
\midrule
\midrule
Pataka & 201 & 82 & 56 & 63 \\
Sischafu & 202 & 82 & 57 & 63 \\
\midrule
\textbf{Subset Repetition} & 201 & 82 & 56 & 63 \\
\midrule
\midrule
Picture Description & 195 & 75 & 54 & 66 \\
Picture Recall & 200 & 81 & 56 & 63 \\
\midrule
\textbf{Subset Picture} & 189 & 74 & 52 & 63 \\
\midrule
\midrule
Story Reading & 199 & 82 & 55 & 62 \\
Story Recall & 185 & 80 & 50 & 55 \\
\midrule
\textbf{Subset Story} & 182 & 80 & 48 & 54 \\
\midrule
\midrule
\textbf{Subset All} & 171 & 73 & 45 & 53 \\
\bottomrule
\end{tabular}
\caption{Number of samples for the PDC data subsets by diagnostic groups: healthy control (HC), mild cognitive impairment (MCI), dementia (DEM).}
\label{tab:subsets}
\end{center}
\end{table}

The audio recordings were collected using a standardized iPad-based application and the same iPad model across all study sites to ensure consistency. 
The recordings include the subjects' speech and, where applicable, brief interventions by the examiner during the performance of the PDTB tasks. 
Participants were given sufficient time to complete the tasks in a quiet, controlled environment.
All iPad data was stored in pseudonymized and encrypted form before being transferred for central processing.
The corpus comprises $\approx$20 hours of raw audio material, with recording durations ranging from 4 seconds to 4.5 minutes, excluding empty or damaged files.
The audio files were recorded as 16-bit mono waveforms with a sampling rate of 44.1~kHz. % and subsequently downsampled to 16 kHz
Since not all subjects completed all of the eight tasks, the corpus provides task-specific subsets (see Table~\ref{tab:subsets}). 

\subsection{Transcription}\label{sc:transcription}
All speech recordings were manually transcribed by a professional transcription service according to the extended scientific transcription rules of Dresing \& Pehl \cite{dresing_pehl10}, preserving verbal and nonverbal features of spoken interaction.

The transcriptions are verbatim, capturing the structure of the spoken language while adapting colloquial speech and dialects to standard German.
For example, reductions such as ``so'n Zettel'' are rendered as ``so ein Zettel,'' and ``hamma'' as ``haben wir.''
% The transcription protocol includes a rich set of paralinguistic and structural markers, which reflect clinically relevant aspects of speech production and conversational organization: Word and sentence breaks (as \texttt{/}), pauses in seconds (e.g., \texttt{(.)}, \texttt{(..)}, \texttt{(...)}, \texttt{(4)}, \texttt{(5)}, ...), fillers and backchannel signals (e.g., ``ähm,'' ``hm,'' ``ja,'' ``aha''), emphasis (as \texttt{VERSALS}), non-verbal utterances (e.g., \texttt{(lacht)}, \texttt{(seufzt)}), incomprehensible utterances (e.g., \texttt{(unv.)}, \texttt{(unv., \textit{with context})}, \texttt{(\textit{guess}?)}), speaker changes (as paragraphs with \texttt{A:}, \texttt{B:} and timestamps), overlapping speech (as \texttt{//}), reception signals with intonation/function (e.g., ``mhm \texttt{(bejahend)},'' ``mhm \texttt{(fragend)}.'')
The transcription protocol includes the following paralinguistic and structural markers reflecting aspects of speech production and conversation organization: 
\begin{itemize}
    \item word and sentence breaks (as \texttt{/})
    \item pauses in seconds (e.g., \texttt{(.)}, \texttt{(..)}, \texttt{(...)}, \texttt{(4)}, \texttt{(5)}, etc.)
    \item filler words and backchannel signals (e.g., ``ähm'', ``hm'', ``ja'', ``aha'')
    \item emphases (as \texttt{VERSALS})
    \item nonverbal utterances (e.g., \texttt{(lacht)}, engl. \texttt{(laughs)})
    \item incomprehensible utterances (e.g., \texttt{(unv.)}, \texttt{(unv., \textit{with context})}, \texttt{(\textit{guess}?)})
    \item speaker changes (as paragraphs beginning with \texttt{A:} or \texttt{B:} and ending with timestamps)
    \item speaker overlaps (as \texttt{//})
    \item reception signals with intonation (e.g., ``mhm \texttt{(bejahend)}'', engl. ``mhm \texttt{(affirmative)}'')
\end{itemize}

The resulting transcripts form the ground truth to which ASR and NLP methods can be measured.
Furthermore, the level of detail enables linguistic, acoustic, and interactional analyses of neurodegenerative speech features such as hesitation, repair phenomena, pause patterns, prosodic intonation, and turn-taking.
A detailed description of the twelve transcription rules and their extensions can be found in \cite{dresing_pehl10}.

\subsection{Metadata}
\begin{table}[!th]
\begin{center}
\begin{tabular}{l|l|l}
\toprule
\textbf{Short}     & \textbf{Education Level}                     & \textbf{Years}       \\
\midrule
\multicolumn{3}{l}{\textbf{education 1: school degree (SD)}} \\
\midrule
SDAL      & A-Level                  & 13          \\
SDPS      & Polytechnical School     & 12          \\
SDSS      & Secondary School         & 10          \\
SDMS      & Main Schooling           & 9           \\
SDWO      & Without                  & 9           \\
SDNC      & No Comment               & 9           \\
\midrule
\multicolumn{3}{l}{\textbf{education 2: degree (D)}}         \\
\midrule
DPHD      & PhD                      & 21          \\
DUNI      & University               & 17          \\
DVTR      & Vocational Training      & = SD + 3    \\
DWTR      & Without Vocational Training & = SD       \\
\bottomrule
\end{tabular}
\caption{Metadata abbreviations for educational degrees 1 and 2 with average years of education.}
\label{tab:ecucation}
\end{center}
\end{table}

During the study visit, a detailed set of metadata was collected for each subject to ensure accurate characterization of the study population and adherence to the study protocol.
General information includes demographic data such as age, gender, and educational level, as well as session parameters such as date, time, state, clinic ID, subject ID, investigator ID, and the iOS version of the iPad used. 
Educational level is reported as average years spent in education and the type of first and second educational degree (see Table~\ref{tab:ecucation}).
Verification of all inclusion and exclusion criteria listed in Section~\ref{sc:criteria} was also documented for each subject.
Diagnosis-related metadata includes the diagnostic group (HC, MCI, DEM) and the Mini Mental State Examination (MMSE) score. 
For MCI and DEM subjects, the NIA-AA criteria for the confirmed AD diagnosis were recorded, including values for the Clinical Dementia Rating (CDR) scale, biomarkers in cerebrospinal fluid ($\beta$-Amyloid and p-Tau), and diagnostic confirmation by imaging biomarkers (AD-related MRI T1/T2/FLAIR or PET).
For more clinical details, refer to the original source of the NIA-AA criteria \cite{clifford2018_niaaa}.

\section{Directions of Research}
The PARLO Dementia Corpus opens a broad range of research opportunities at the intersection of speech technology, linguistics, and clinical neuroscience.
Its multi-center design, standardized task structure, and detailed clinical annotations make it a valuable resource for computer-based and clinical investigations of cognitive decline.

From a speech technology perspective, the corpus provides a benchmark for the development and evaluation of ASR systems for disordered and atypical speech. 
The recordings capture variations in speech fluency, articulation, and lexical access typical of MCI and dementia, enabling studies on robustness, domain adaptation, and bias mitigation in multilingual ASR systems.
Beyond ASR, the data enable the extraction of acoustic (e.g., pause duration, prosodic variability, and voice quality) and linguistic biomarkers (e.g., lexical, syntactic, and semantic changes) associated with the progression of neurodegenerative diseases.
The structured form of the PDTB allows for a systematic comparison of speech and language production under different cognitive loads, ranging from constrained naming to phonological production to open-ended narrative recall. 
This diversity enables the investigation of task-dependent and cross-task features, as well as multi-modal modeling approaches that combine text and audio signals.
%Its multi-center design also allows for systematic studies on dialectal variations and recording conditions. 

From a clinical perspective, the corpus provides an empirical basis for the development and validation of automated screening tools for the early detection and monitoring of AD. 
By linking speech-based measurements with established clinical scales (e.g., MMSE) and biomarkers (e.g., cerebrospinal fluid), the dataset supports clinically interpretable modeling of cognitive impairments.

Beyond computational approaches, the PDC establishes a benchmark for speech-based dementia research in the German language. 
Its compatibility with established English corpora such as Pitt Corpus supports cross-lingual comparisons and contributes to the broader goal of developing reliable speech technologies for early cognitive screening.

To ensure the comparability and reproducibility of future studies using the PDC, we recommend standardized evaluation procedures that account for speaker (and site) variability, such as speaker-independent cross-validation (and leave-one-site-out validation).

\section{Baseline Experimental Design}
\subsection{Automatic Speech Recognition}\label{sc:asr}
We automatically transcribe all speech recordings in the PDC using three state-of-the-art ASR systems: \texttt{whisper-large-v3} by OpenAI \cite{whisper}, \texttt{parakeet-tdt-0.6b-v3} by NVIDIA, and \texttt{owsm\_ctc\_v4\_1B} by Peng et al. \cite{peng25c_interspeech}.
All models are openly available from HuggingFace\footnote{\url{www.hf.co/openai/whisper-large-v3}}\footnote{\url{www.hf.co/nvidia/parakeet-tdt-0.6b-v3}}\footnote{\url{www.hf.co/espnet/owsm_ctc_v4_1B}}.
The full raw audio recordings of the PDTB tasks (see Section~\ref{sc:tasks}) are transcribed with each ASR system using the default decoding parameters and the German language set. 
The word repetition tasks (i.e., ``pataka'' and ``sischafu'') are excluded from ASR evaluation because their output requires phonetic rather than orthographic evaluation.
Transcripts are normalized regarding case-sensitivity, punctuation, fillers, and numerical tokens prior to evaluation.

\subsection{Automatic Test Evaluation}\label{sc:scoring}
For the automatic evaluation of the standardized cognitive tests included in the PDTB, rule-based scoring is applied to the automatic transcriptions of the \textit{Confrontation Naming} and \textit{Animal Naming} tests. 
In the \textit{Confrontation Naming} test, performance is evaluated as the number of correctly named objects among the 15 presented stimuli, while in the \textit{Animal Naming} test, the score corresponds to the number of uniquely named animals within one minute. 
To compute these scores automatically, the ASR transcripts are normalized and compared against predefined task-specific lexical sets. 
The reference lists comprise the target object names (\textit{Confrontation Naming}) and animal lexicon (\textit{Animal Naming}), which are derived from GermaNet \cite{hamp-feldweg-1997-germanet, henrich-hinrichs-2010-gernedit} and the German OpenThesaurus\footnote{\url{www.openthesaurus.de}} for synonym and morphological variant matching. 
Word matches are identified using Levenshtein distance with a minimum threshold of 85\% relative to word length. 
After filtering out duplicates, the word count represents the final score.

\subsection{Large Language Models}\label{sc:llm}
For language modeling and classification experiments, we employ the open-source Vision Large Language Model (Vision-LLM) \texttt{Mistral-3.1}\footnote{\url{www.hf.co/RedHatAI/Mistral-Small-3.1-24B-Instruct-2503-FP8-dynamic}}, which is based on the Mistral 24B architecture.                              
To enable efficient inference on a single GPU (\texttt{NVIDIA L40S, 48 GB VRAM}), the model is deployed using FP8 quantization and executed via the vLLM inference engine \cite{kwon2023efficient}. 
Although FP16 inference can yield slightly higher numerical precision, FP8 was chosen to balance computational efficiency and model usability.
Model generation follows the parameters recommended by the authors (\texttt{temperature=0.15}, \texttt{max\_tokens=4096}) to ensure reproducibility and prevent excessive output generation. 
To ensure consistent and structured output, all completions use constrained decoding based on a predefined JSON schema that specifies the output format. 
All prompts are written in German to match the input language, while English translations are provided in this article.
A consistent system prompt is applied across all experiments:
\begin{tcolorbox}[colframe=black!60, colback=gray!10, boxrule=0.5pt, width=\columnwidth, boxsep=1pt,  left=2pt, right=2pt, top=2pt, bottom=2pt]
\textit{``You are an expert in Alzheimer's dementia who classifies the cognitive status of individuals based on their speech performance. You must return a JSON object that strictly follows this schema:''}  \\ + JSON schema
\end{tcolorbox}
This is followed by a task-specific user prompt containing a brief description of the cognitive task (e.g., picture description), the definition of the classification target (HC vs. MCI vs. DEM), and the corresponding transcription input from the subject's task execution. 
We additionally input the source picture (Figure~\ref{fig:mountain}) for the \textit{Picture Description} task and the ground truth text for the \textit{Story Reading} task.
Original prompts are given in the supplementary material \ref{sc:prompts}.

\section{Results}
\subsection{ASR Performance}
\begin{table}[!th]
\begin{center}
\begin{tabular}{l|l||c|c|c}
\toprule
\textbf{Task}                                                                            & \textbf{Group} & \textbf{whisper} & \textbf{OWSM}  & \textbf{parakeet} \\ \midrule\midrule
\multirow{4}{*}{\begin{tabular}[c]{@{}l@{}}\textbf{Story} \\ \textbf{Reading}\end{tabular}}       & HC   & 5.0     & 9.7   & 6.8      \\
                                                                                & MCI   & 8.2     & 14.2  & 9.4      \\
                                                                                & DEM   & 15.2    & 19.8  & 14.0     \\ \cmidrule{2-5}
                                                                                & ALL   & 9.0     & 14.1  & 9.7      \\ \midrule\midrule
\multirow{4}{*}{\begin{tabular}[c]{@{}l@{}}\textbf{Confron-} \\ \textbf{tation} \\ \textbf{Naming}\end{tabular}}       & HC   & 10.9    & 26.8  & 11.1     \\
                                                                                & MCI   & 20.2    & 35.3  & 20.7     \\
                                                                                & DEM   & 23.6    & 38.5  & 23.8     \\ \cmidrule{2-5}
                                                                                & ALL   & 17.6    & 32.9  & 17.9     \\ \midrule\midrule
\multirow{4}{*}{\begin{tabular}[c]{@{}l@{}}\textbf{Animal} \\ \textbf{Naming}\end{tabular}}       & HC   & 13.9    & 35.4  & 20.7     \\
                                                                                & MCI   & 21.6    & 39.8  & 22.9     \\
                                                                                & DEM   & 18.9    & 38.8  & 24.2     \\ \cmidrule{2-5}
                                                                                & ALL   & 17.7    & 37.7  & 22.4     \\ \midrule\midrule
\multirow{4}{*}{\begin{tabular}[c]{@{}l@{}}\textbf{Picture} \\ \textbf{Descrip-} \\ \textbf{tion}\end{tabular}} & HC   & 7.6     & 13.4  & 7.8      \\
                                                                                & MCI   & 10.5    & 15.9  & 10.4     \\
                                                                                & DEM   & 18.5    & 26.4  & 17.1     \\ \cmidrule{2-5}
                                                                                & ALL   & 12.1    & 18.5  & 11.7     \\ \midrule\midrule
% \multirow{4}{*}{pataka}                                                         & HC   & 23.9    & 65.2  & 61.5     \\
%                                                                                 & MCI   & 26.8    & 70.1  & 59.2     \\
%                                                                                 & DEM   & 37.3    & 78.7  & 62.1     \\ \cmidrule{2-5}
%                                                                                 & ALL   & 28.9    & 70.8  & 61.0     \\ \midrule\midrule
% \multirow{4}{*}{shishafu}                                                       & HC   & 166.0   & 121.5 & 127.9    \\
%                                                                                 & MCI   & 159.6   & 114.8 & 142.7    \\
%                                                                                 & DEM   & 149.4   & 110.9 & 140.2    \\ \cmidrule{2-5}
%                                                                                 & ALL   & 159.0   & 116.3 & 135.9    \\ \midrule\midrule
\multirow{4}{*}{\begin{tabular}[c]{@{}l@{}}\textbf{Story} \\ \textbf{Recall}\end{tabular}}        & HC   & 10.8    & 15.5  & 11.7     \\
                                                                                & MCI   & 17.7    & 23.3  & 19.1     \\
                                                                                & DEM   & 25.2    & 35.8  & 37.4     \\ \cmidrule{2-5}
                                                                                & ALL   & 16.9    & 23.7  & 21.4     \\ \midrule\midrule
\multirow{4}{*}{\begin{tabular}[c]{@{}l@{}}\textbf{Picture} \\ \textbf{Recall}\end{tabular}}      & HC   & 8.4     & 12.1  & 10.9     \\
                                                                                & MCI   & 15.1    & 23.8  & 17.5     \\
                                                                                & DEM   & 16.9    & 28.8  & 21.5     \\ \cmidrule{2-5}
                                                                                & ALL   & 12.9    & 20.6  & 16.1     \\ 
\bottomrule
\end{tabular}
\caption{Average Word Error Rate (in \%) by diagnostic group (HC: healthy hontrol, MCI: mild cognitive impairment, DEM: dementia) and cognitive task for \texttt{whisper-large-v3}, \texttt{parakeet-tdt-0.6b-v3}, and \texttt{owsm\_ctc\_v4\_1B}.}
\label{tab:wer}
\end{center}
\vspace{-0.6em}
\end{table}

Automatic transcription quality for the ASR models described in Section~\ref{sc:asr} is evaluated based on the word error rate (WER) relative to normalized human reference transcriptions (see Section~\ref{sc:transcription}). 
The average WER (in \%) is calculated per cognitive task and diagnostic group using normalized transcripts from \texttt{whisper-large-v3} (\texttt{whisper}), \texttt{parakeet-tdt-0.6b-v3} (\texttt{parakeet}), and \texttt{owsm\_ctc\_v4\_1B} (\texttt{OWSM}), respectively. 

The ASR results are summarized in Table~\ref{tab:wer}.
Across all tasks, \texttt{whisper} consistently achieves the lowest WER, followed by \texttt{parakeet} and \texttt{OWSM}.
On average, \texttt{whisper} reaches an overall WER of 9--18\% depending on the task, substantially outperforming the other models (typically 5--10 pp lower).
Performance differences between the diagnostic groups are consistent across all models, with error rates increasing from healthy controls (HC) to mild cognitive impairment (MCI) to dementia (DEM). 
This trend reflects the increasing prevalence of speech disorders, pauses, and articulation irregularities in cognitively impaired speakers, which pose challenges for current end-to-end ASR systems.
The task-specific speech production context also strongly influences transcription accuracy. 
The lowest WERs are obtained for \textit{Story Reading} and \textit{Picture Description}, which involve structured and continuous speech. 
In contrast, spontaneous and memory-intensive tasks such as \textit{Story Recall} and \textit{Animal Naming} show markedly higher WERs (up to 25\% for DEM).
These increased error rates are partly attributable to the intrinsic task characteristics: Subjects with cognitive impairment often produce shorter, fragmentary responses, hesitations, and prolonged pauses, leading to higher WERs. 
\texttt{OWSM} exhibits the largest degradation under these conditions, suggesting lower robustness to non-native and disordered speech compared to \texttt{whisper}'s large multilingual training data coverage.

\subsection{Automatic Test Evaluation}
These experiments aim to automatically evaluate the standardized subtests of the PDTB, which are the \textit{Confrontation Naming} and the \textit{Animal Naming} tasks, by applying the rule-based scoring approach from Section~\ref{sc:scoring}. 
We use transcriptions generated by \texttt{whisper}, which achieved the best ASR performance for both tasks (see Table~\ref{tab:wer}).
Scoring performance is measured using the Pearson correlation coefficient between the automatic scores derived from ASR transcripts and those derived from human reference transcripts. 
In addition, the calculated test scores are correlated with the expert-assigned MMSE scores to verify their diagnostic validity.

The results show nearly perfect correlations between ASR-based and reference-based scores, reaching a value of 0.92 for \textit{Confrontation Naming} and 0.99 for \textit{Animal Naming}. 
These findings confirm that automatic scoring based on ASR transcripts can reproduce the manual evaluation outcomes. 
As expected, the correlations with global MMSE scores are lower, at 0.43 for \textit{Confrontation Naming} and 0.65 for \textit{Animal Naming}.
However, the combination (addition) of both subtest scores achieves a correlation of 0.70 with the MMSE, suggesting that the automatic evaluation of combined verbal tasks can capture clinically relevant deviations with global scales.

\subsection{LLM Zero-Shot Classification}
\begin{figure}[!th]
  \begin{center}
  \includegraphics[width=\columnwidth]{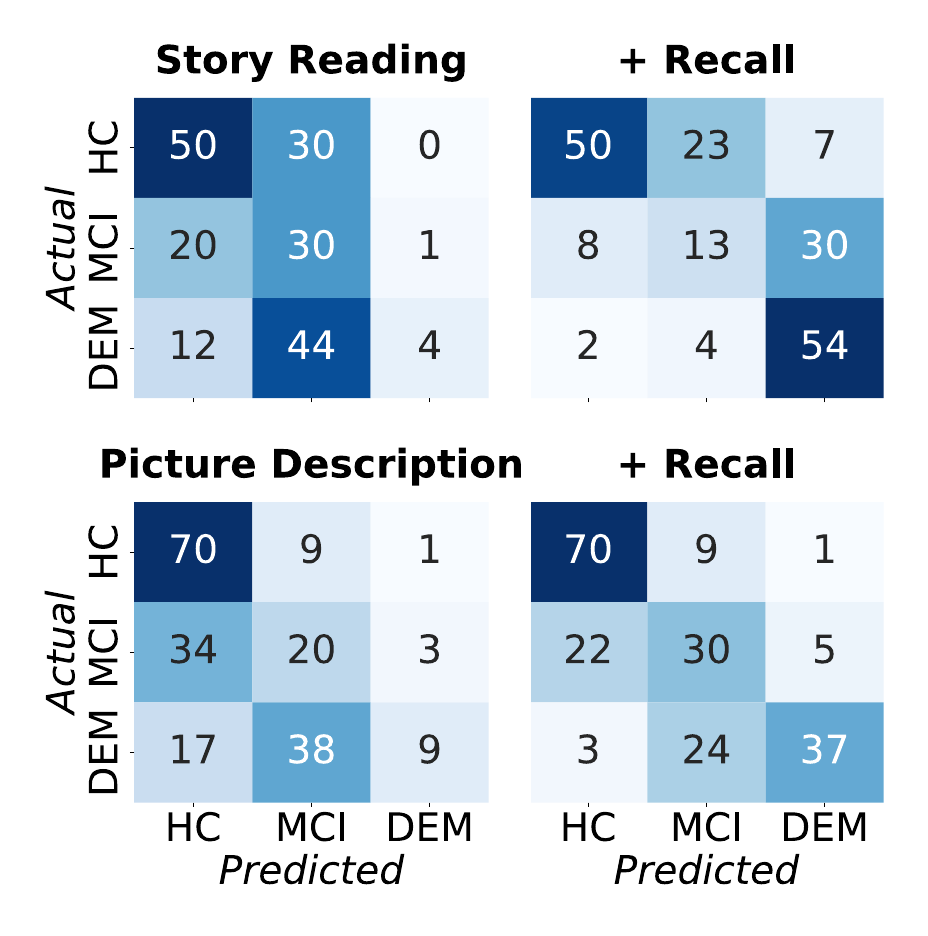}
  \caption{Confusion matrices for LLM zero-shot classification of healthy control (HC), mild cognitive impairment (MCI), and dementia (DEM) groups.}
  \label{fig:confmat}
  \end{center}
\end{figure}
Our classification experiments aim to evaluate the ability of a Vision-LLM to determine the cognitive status of individuals in a zero-shot setting without task-specific fine-tuning. 
The objective is a 3-class classification that distinguishes between healthy controls (HC), and individuals with AD-related mild cognitive impairment (MCI) and dementia (DEM). 
The speech samples are taken from the \textit{Story} and \textit{Picture} subsets of the corpus (see Table~\ref{tab:subsets}), including both stimulus tasks (\textit{Story Reading}, \textit{Picture Description}) and the corresponding recall tasks (\textit{Story Recall}, \textit{Picture Recall}). 
Automatic transcriptions are generated using \texttt{whisper}, which achieved the best overall ASR performance in the tasks (Table~\ref{tab:wer}). 
Prompts and system configurations correspond to the settings described in Section~\ref{sc:llm}. 
Two variants are tested: (1) baseline prompts using only the stimulus task, and (2) extended prompts that additionally include the corresponding recall transcription.
Model performance is evaluated using the Unweighted Average Recall (UAR) to ensure equal weighting of all diagnostic classes. 

Results are shown as confusion matrices in Figure~\ref{fig:confmat}. 
For both the story tasks and the picture tasks, the UAR increases significantly from  42.7\% and 45.6\% (stimulus only) to 59.3\% and  66.0\% when recall speech is included.
These improvements indicate that incorporating recall data enhances discriminative performance, particularly by improving the model's sensitivity to intermediate impairment (MCI). 
As shown in Figure~\ref{fig:confmat}, MCI samples were frequently confused with neighboring classes in the baseline setting and misclassified as HC in \textit{Story Reading} and as DEM in \textit{Picture Description}. 
This may be because \textit{Picture Description} is more cognitively demanding than \textit{Story Reading}.
Adding recall inputs reduces these confusions and yields a more balanced class distribution, reflecting the diagnostic significance of memory-driven, spontaneous speech for cognitive differentiation.

\section{Conclusion}
We introduced the PARLO Dementia Corpus (PDC), the first multi-center, clinically validated German resource addressing AD-related mild cognitive impairment and dementia. 
The corpus includes audio recordings of AD patients and healthy controls performing eight established neuropsychological tasks covering speech and language production under various cognitive loads, from constrained naming to phonological production to open-ended narrative recall. 
Enriched with detailed demographic, clinical and biomarker metadata, as well as human-verified transcriptions, the PDC provides a comprehensive resource for research into speech and cognition in neurodegenerative diseases. 
With its standardized data collection protocol and open evaluation framework, the corpus supports a wide range of research directions, including automatic speech recognition for pathological and atypical speech, extraction of acoustic and linguistic biomarkers, multi-modal modeling (audio, text, medical biomarker), and automated cognitive assessment. 
Initial benchmark experiments demonstrate the feasibility of LLM-based zero-shot classification of cognitive impairment groups and the automatic evaluation of standardized dementia tests, providing a basis for future research.

\section{Data Availability and Licensing}
Researchers interested in obtaining access to the PDC are invited to contact PARLO at \texttt{info@parlo-institut.de} for commercial and non-commercial licensing. 
Requests will be reviewed in accordance with ethical and data protection requirements.

% \section{Acknowledgements}

% Place all acknowledgments (including those concerning research grants and funding) in a separate section at the end of the paper.

\section{Bibliographical References}\label{sec:reference}
\bibliographystyle{lrec2026-natbib}
\bibliography{main}

\section{Ethics Statement}
All data included in the PARLO Dementia Corpus were collected in compliance with national and institutional ethical standards for medical research involving human participants. 
The study protocol was approved by the responsible IRBs at all participating clinical centers in Germany, and all participants (or their legal representatives) provided written informed consent prior to inclusion. 
Participants were informed about the study's aims, the nature of the speech recordings, and their right to withdraw consent at any time without negative consequences. 
All recordings were pseudonymized at the point of collection, and no personally identifiable information is contained in the released dataset. 
Data transfer, storage, and processing followed the principles of the EU General Data Protection Regulation. 
Access to the corpus requires a signed data-use and license agreement ensuring compliance with ethical and legal standards. 
The PARLO Dementia Corpus has been curated with particular attention to participant dignity, data security, and scientific transparency, aiming to enable reproducible and ethically responsible research on speech and cognitive health.

\appendix
\section{Appendix}
\subsection{Prompts}\label{sc:prompts}
The following prompts (in German to match the input language) are used in the LLM zero-shot classification experiments from Section~\ref{sc:llm}.

A unified system prompt with JSON schema:
\begin{tcolorbox}[colframe=black!60, colback=gray!10, boxrule=0.5pt, width=\columnwidth, boxsep=1pt,  left=2pt, right=2pt, top=2pt, bottom=2pt]
\textit{``Sie sind ein Experte für Alzheimer's Demenz, der den kognitiven Status von Personen anhand ihrer Sprachleistung klassifiziert.
    \\Sie müssen ein JSON-Objekt zurückgeben, das sich strikt an dieses Schema hält:
    \\\texttt{\{'properties': 
    \\\{'cognitive\_status': 
    \\\{'enum': ['HC', 'MCI', 'DEM'], 
    \\'title': 'Cognitive Status', 'type': 'string'\}\}, 
    \\'required': ['cognitive\_status'], 
    \\'title': 'CognitiveAssessment', 
    \\'type': 'object'\}}
    %{'properties': {'cognitive_status': {'enum': ['HC', 'MCI', 'DEM'], 'title': 'Cognitive Status', 'type': 'string'}}, 'required': ['cognitive_status'], 'title': 'CognitiveAssessment', 'type': 'object'}
    ''}
\end{tcolorbox}

A prompt for the \textit{Picture Description} task with the extension for \textit{Picture Recall} task (in parentheses):
\begin{tcolorbox}[colframe=black!60, colback=gray!10, boxrule=0.5pt, width=\columnwidth, boxsep=1pt,  left=2pt, right=2pt, top=2pt, bottom=2pt]
\textit{``Die Person beschreibt das gegebene Bild (Bildbeschreibung).
    \\Nach Ablenkung muss die Person das Bild aus dem Gedächtnis beschreiben (Verzögerter Recall).
    \\Das Bild kann mit acht Konzepten beschrieben werden:
    \\- Eine Berglandschaft
    \\- Wanderer suchen den Weg
    \\- Eine Brücke führt über den Fluss
    \\- Kinder sitzen auf einem Baumstamm
    \\- Ein Junge ruft seine Eltern
    \\- Ein Mädchen fällt ins Wasser
    \\- Gewitterwolken mit Blitzen
    \\- Ein Haus in der Ferne
    \\
    \\Ihre Aufgabe darin, anhand der Bildbeschreibung und des verzögerten Recalls den kognitiven Status der Person in eine der drei Kategorien zu klassifizieren:
    \\- HC = Healthy Control
    \\- MCI = Mild Cognitive Impairment
    \\- DEM = Mild to Severe Dementia
    \\
    \\Hier ist die Bildbeschreibung:
    \\... \texttt{\{picture\_description\}} ...
    \\
    \\(Hier ist der verzögerte Recall:)
    \\(... \texttt{\{picture\_recall\}} ...)
    \\
    \\Geben Sie die Bewertung NUR im JSON-Format entsprechend des gegebenen Schemas an. Keine anderen Ausgaben.''}
\end{tcolorbox}

A prompt for the \textit{Story Reading} task with the extension for \textit{Story Recall} task (in parentheses):
\begin{tcolorbox}[colframe=black!60, colback=gray!10, boxrule=0.5pt, width=\columnwidth, boxsep=1pt,  left=2pt, right=2pt, top=2pt, bottom=2pt]
\textit{``Die Person liest eine Geschichte vor (Vorlesung).
    \\Nach Ablenkung erzählt die Person die Geschichte aus dem Gedächtnis mit eigenen Worten nach (Verzögerter Recall).
    \\
    \\Ihre Aufgabe besteht darin, anhand der Vorlesung und des verzögerten Recalls den kognitiven Status der Person in eine der drei Kategorien zu klassifizieren:
    \\- HC = Healthy Control
    \\- MCI = Mild Cognitive Impairment
    \\- DEM = Mild to Severe Dementia
    \\
    \\Hier ist die Vorlesung:
    \\... \texttt{\{story\_reading\}} ...
    \\
    \\(Hier ist der verzögerte Recall:)
    \\(... \texttt{\{story\_recall\}} ...)
    \\
    \\Geben Sie die Bewertung NUR im JSON-Format entsprechend des gegebenen Schemas an. Keine anderen Ausgaben.''}
\end{tcolorbox}
% \section{Language Resource References}
% \label{lr:ref}
% \bibliographystylelanguageresource{lrec2026-natbib}
% \bibliographylanguageresource{languageresource}

\end{document}